\documentclass{article}

\usepackage{arxiv}

\usepackage{amsfonts,amsmath}
\usepackage{amssymb,amsbsy}
\usepackage{soul,color}

\usepackage[utf8]{inputenc} 
\usepackage[T1]{fontenc}    
\usepackage{hyperref}       
\usepackage{url}            
\usepackage{booktabs}       
\usepackage{amsfonts}       
\usepackage{nicefrac}       
\usepackage{microtype}      
\usepackage{cleveref}       
\usepackage{graphicx}
\usepackage{natbib}
\usepackage{doi}

\title{Graded damage solutions in one dimension}

\date{\today}

\author{ \href{https://orcid.org/0000-0001-8830-1964}{\includegraphics[scale=0.06]{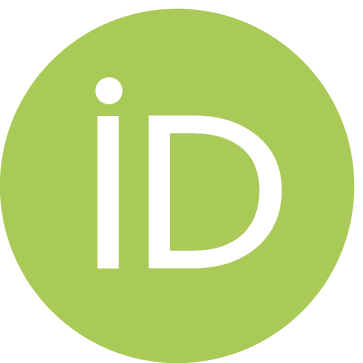}\hspace{1mm}Nunziante Valoroso}\\
	Dipartimento di Ingegneria\\
	Università di Napoli Parthenope\\
	Napoli, 80143 \\
	\texttt{nunziante.valoroso@uniparthenope.it} \\
}

\renewcommand{\shorttitle}{Graded damage solutions}

\hypersetup{
pdftitle={Graded damage solutions},
pdfsubject={math-ph},
pdfauthor={Nunziante Valoroso},
pdfkeywords={Damage mechanics, Regularization, Cohesive zone models},
}

\newcommand{\dom}{d}
\newcommand{\dam}{\dom}
\newcommand{\ddam}{\dot{\dam}}

\newcommand{\elib}{\psi}
\newcommand{\elibint}{\tilde\psi}

\newcommand{\eps}{\boldsymbol{\varepsilon}}

\newcommand{\deri}[2]{\dfrac{\partial #1}{\partial #2}}

\newcommand{\sig}{\sigma}

\newcommand{\dd}{\;\mathrm{d}}

\newcommand{\inv}[1]{\dfrac{1}{#1}}
\newcommand{\Epot}{\mathcal{E}}
\newcommand{\Diss}{\mathcal{D}}

\newcommand{\demi}{\inv{2}}

\newcommand{\be}[1]{\begin{equation}
#1
\end{equation}}

\newcommand{\ddom}{\dot{\dom}}
\newcommand{\deldom}{\delta\dom}

\newcommand{\domstar}{\dom^*}

\newcommand{\Young}{E}

\newcommand{\lc}{l_c}

\def\leftj{\left\lbrack\!\!\left\lbrack}
\def\rightj{\right\rbrack\!\!\right\rbrack}

\newcommand{\udepl}{{u}}

\newcommand{\dx}{\dd x}
\newcommand{\Yc}{Y_c\,}
\newcommand{\Ycd}{Y_c(\dom)}
\newcommand{\Ycpd}{Y^\prime_c(\dom)\,}
\newcommand{\Yczero}{Y_c(0)\,}

\newcommand{\Y}{Y}

\newcommand{\Yint}{\tilde{Y}}

\newcommand{\Ycint}{\tilde{Y}_c\,}
\newcommand{\Ycdint}{\tilde{Y}_c(\dom)}

\newcommand{\dm}{\dom_m}

\newcommand{\ustar}{u^\star}

\newcommand{\gdue}{g_2(\dom)}

\newcommand{\Hm}{H(\dom_m)}
\newcommand{\Fm}{F(\dom_m)}
\newcommand{\uel}{u_{el}}
\newcommand{\omegad}{\omega}
\newcommand{\lm}{l_m}
\newcommand{\wcoh}{w}
\newcommand{\lcoh}{l_{coh}}
\newcommand{\sigmaf}{\sigma_c}
\newcommand{\sigmafq}{\sigma^2_c}
\newcommand{\Gf}{G_c}

\newcommand{\Gzero}{G_0}
\newcommand{\Gc}{G_c}

\newcommand{\intLL}{\int_\Omega}
\newcommand{\intL}{\int_\Omega}

\newcommand{\sign}{\mathrm{sign}}

\newcommand{\Auno}{A_1}
\newcommand{\Adue}{A_2}

\newcommand{\udeplprime}{\frac{\dd\udepl}{\dx}}

\newcommand{\damprime}{\frac{\dd\dam}{\dx}}

\newcommand\unmezzo{\frac{1}{2}}

\newcommand\epsint{{w}}
\newcommand\epsintq{{\epsint^2}}

\newcommand\epsintzero{{\epsint_0}}

\newcommand\epsintc{{\epsint_c}}

\newcommand\sigint{t}

\newcommand\kk{k\,}

\newcommand\Preac{P}

\begin{document}
\maketitle

\begin{abstract}
A regularized damage model is considered
named \emph{Graded damage} in which the gradient enhancement has
the form of an explicit bound for the spatial gradient of damage.
The key features of the proposed approach are demonstrated
by computing the analytical solution of two problems
that are one-parameter dependent.
The first one is the classical one-dimensional damageable rod
under tensile load, for which the hardening
function is determined based on the
equivalence with a given cohesive relationship.
The second application is a mode-I
delamination problem for which the cohesive law for the
interface is formulated starting
from the graded damage concept, i.e.~by prescribing the shape of damage
distribution within the cohesive process zone.
\end{abstract}



\section{Introduction}
\label{sec1}

Damage and Fracture Mechanics find their \emph{raison d'\^{e}tre}
in the need for predictive computations able to
prevent catastrophic failure in engineering structures.
The complexity of the physics of damage,
which rules out any homogeneity
of materials at the usual macroscopic scale
of laboratory experiments,
has led to many different modeling assumptions
in Solid Mechanics, each of them
resulting from a suitable trade-off
between physical relevance at different scales
and applicability to structural design \cite{BazantPlanas:98}.
Nonetheless, computations of failure mechanisms and
ultimate load-carrying capacity of structures
still stay as difficult tasks
in civil and mechanical engineering
owing to the intrinsic non-smoothness of
damage and fracture phenomena \cite{Zaoui:2013}.

Roughly speaking, one can
categorize the computational approaches to failure
into two families, the continuous and the discontinuous one,
each of them with advantages and limitations.
Discontinuous descriptions allow for jumps
in the displacement field, whereby one has to
deal with changes in topology that
are intrinsic to the representation of discrete cracks.
In a finite element context this requires
special elements with embedded discontinuities \cite{Armero:1996}
or extended finite element formulations (X-FEM)
either in the original setup of Belytschko and co-workers
\cite{Moes:1999} or in the format of the so-called \emph{Thick Level Set}
model \cite{Moes:2011}.

One can also include into the discontinuous family
the cohesive zone models originating
from the work of Barenblatt \cite{Barenblatt:1962}.
Initially motivated by the need to characterize
stress states in the vicinity of cracks,
in cohesive zone models one may speak e.g.~of damage, delamination
or de-cohesion to designate all those progressive phenomena
preceding fracture, during which material separation
is resisted by attractive forces
that develop along an extended crack tip,
i.e.~the \emph{cohesive process zone}.

In the cohesive zone approach
crack progression is governed by an independent
relationship between surface
tractions and displacement jumps
that incorporates typical fracture parameters,
i.e.~the cohesive strength and the fracture toughness.
Under certain conditions the shape of the softening curve
does also play a role in fracture predictions \cite{Alfano:2006},
but it is commonly believed
to be less relevant compared to the other parameters.
Anyway, classical implementations of the cohesive zone concept only allow
for strong discontinuities along interfaces
that pre-exist in the material before any loading,
whereby in numerics use is made of
degenerated (zero-thickness) finite elements
that are
placed along potential discontinuity surfaces \cite{Mi98}.

A discrete crack representation
closely reflects the physics of fracture
but includes a number of difficulties,
most of which are related to crack tracking.
This partly motivates
the continuous approach as a tool
for modeling fracture starting from the
strain localization stage;
here no physical crack opening
exist and fully damaged states
can be understood as the smeared, diffuse
representation of macro-fractures.
In this context the basic idea consists of
preserving the topology of the
initial finite element mesh and
to bring into the material model
a concise information about material microstructure
via a length scale parameter;
the latter is used to
introduce the necessary regularization
that restores well-posedness of the problem
either via
a nonlocal integral approach after
Pijaudier-Cabot and Ba{\v{z}}ant \cite{Pijaudier:1987}
or in the form of a gradient enhancement
in the wake of the works of Peerlings et al.
\cite{Geers1998133,Peerlings:1996}.
We also note the family of
\emph{phase-field models} initiated
from the regularized form of the variational theory
of quasi-static fracture \cite{Bourdin:2000};
though starting from a different perspective,
i.e.~global energy minimization,
these models end up with a field equation of diffusive type
that is quite close to that of gradient damage models,
see e.g.~\cite{Miehe:2010b,Miehe:2010a} among others.

In all such cases averaging or differential operators come into play,
whereby the constitutive equations are
no longer defined at the local level but
are established at the scale of the structural model.
One may then conclude that continuous representations of
discontinuities provide globally
smoothed solutions through elements, whereas
in usual local models
stresses, strains and internal variables are all defined
in a point-wise fashion
that can be understood as generally discontinuous fields
inside elements and across elements boundaries \cite{Alfano1998325}.


This chapter is concerned with
a gradient-based continuum damage formulation
named \emph{Graded damage}
\cite{Valoroso-Stolz:2022},
i.e.~a Generalized Standard Model with convex constraints
that admits the geometrical interpretation of the
Thick Level Set approach of Mo\"{e}s et al.
\cite{Moes:2011}.
The variational structure of the model along with its
directional convexity properties
allow for an effective implementation based
on convex programming in the spirit of \emph{direct methods},
by alternating minimization with respect to displacements and damage and maximization with respect
to the Lagrange multipliers that
implicitly contain the information necessary
to track the interphases between
fully damaged regions and the sound material.

In particular, in the following the graded damage model is applied to
two different one-dimensional problems
for which a non-homogeneous solution is computed in closed form.
The first problem is presented in Section \ref{sec2};
it is rather classical and refers to the
rod under tensile load,
whose interest lies in the fact that the relevant solution
is considered to be representative of the response of a
three-dimensional structure across a localization band.
Moreover, this analytical solution is typically used to design the
constitutive functions of a continuum damage formulation
that render the response of the damageable rod
identical to the one of an elastic bar
in which a cohesive interface is the only source of dissipation,
see e.g. \cite{LorentzGodard:2011,Wu:2017}.

The second problem is discussed in Section \ref{sec3};
it is in a sense
analogous to the one of the tensile rod
and refers to a one-parameter-dependent delamination problem
where the interface constitutive relationship
is gradient-enhanced
based on the graded damage concept.
Worth noting is the fact that
in cohesive models understood in the sense
of Hillerborg \cite{Hillerborg76}
there is in principle no need for any regularization,
whereby few attempts have been made so far to
introduce gradients along a cohesive interface.
However, the general consensus that nonlocal
interactions may occur at the meso-scale level
suggest that introduction of a material length scale into an
interface model is a worthwhile attempt
\cite{Latifi:2017,NGUYEN:2016}.


\section{The damageable rod}
\label{sec2}

\begin{figure}[hbt!]
\begin{center}
\includegraphics[width=0.9\textwidth]{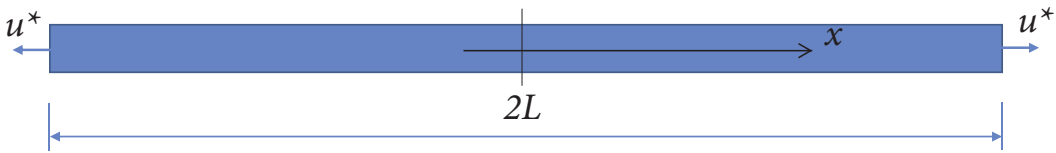}
\end{center}
\caption{One-dimensional rod. Model problem.}
\label{fig-oned-1}
\end{figure}

To begin with, consider the
homogeneous elastic-damageable rod
depicted in Figure \ref{fig-oned-1};
the domain $\Omega$ occupied by the structure is the interval $[-L, L]$,
body forces are neglected and loading is performed
via an increasing elongation at the two ends of the bar.
%

In the present one-dimensional context
the stored energy function
reads:
\be{
\elib(\udepl,\dam)=\demi\,\omegad(\dom)\,E\,\left(\udeplprime\right)^2
\label{eq-freeenergy}}
where $\Young>0$ is the elastic modulus,
$\udepl$ is the axial displacement
obeying the essential boundary conditions:
\be{ \udepl(-L)=-\ustar,\quad \udepl(L)=\ustar; \qquad \ustar>0}
$\dom$ is the damage variable and
$\omegad(\dom)$ is a monotonically
decreasing function
that accounts for material degradation.

Restricting attention to the class of
Generalized Standard Materials,
the local model is completed by prescribing a
dissipation pseudo-potential \cite{Halphen:1975};
for rate-independency,
it must be positively homogeneous of degree-one
with respect to the flux $\ddam$:
 \begin{equation}
\varphi(\ddam)=\Ycd\,\ddam+\sqcup_{\Re^+}(\ddam)
\label{eq-localdiss}
\end{equation}

In the above relationship $\Ycd$ is a positive convex function
of the current damage state $\dom$,
here considered as a parameter,
and $\sqcup_{\Re^+}$ is the convex indicator
of non-negative reals
that enforces irreversibility:
  \begin{equation}
\sqcup_{\Re^+}(\domstar)=
\left\{
\begin{array}{ll}
  0\qquad &{\rm{if}}\,\domstar\geq 0\\
  \noalign{\medskip}
  +\infty &{\rm{otherwise}}
\end{array}
\right.
  \label{damageRplusindicator}
  \end{equation}

Likewise, damage is a constrained variable since it
has to comply with the physical bounds:
\begin{equation}
0\leq\dam\leq1
\label{eq-d-bounds}
\end{equation}
%
This can be accounted for in the present formulation
either via an indicator function of the admissibility domain
or using a smoothed version of it, say $g_1(\dom)$,
with the relevant Karush-Kuhn-Tucker conditions
\begin{equation}
g_1(\dom)\leq 0; \qquad \gamma_1\geq0; \qquad \gamma_1\,g_1(\dom)=0
\label{liag1}
\end{equation}

%

It is well known that
the local constitutive equations emanating from the
above potentials will produce non-objective numerical solutions
with respect to finite element meshes;
actually, owing to strain softening,
strains and damage do localise into narrow regions
with high gradients and
mechanical dissipation is
strongly affected by mesh refinements.
Objectivity can be restored by appealing to
a nonlocal formulation, i.e.~introducing
spatial interactions into
the constitutive equations
to provide a suitable localization limiter \cite{BazantPlanas:98}.
To this end, in the graded damage model
an explicit nonlocal constraint acting
on the damage gradient is prescribed
via the following \cite{Valoroso-Stolz:2022}:
 \be{ \gdue=\left\vert\vert\nabla\dom\right\vert\vert-f(\dom)  
 \label{oned-constraint}}
%
where the bounding function $f(\dom)>0$
may be arbitrarily nonlinear provided
that it is concave; in addition,
the following complementarity conditions apply:
\be{
\gdue\leq0\;; \qquad \gamma_2\geq 0\;; \quad \gamma_2\,\gdue=0
\label{oned-KTconditions}}
that characterize the gradient constraint
(\ref{oned-constraint}) as non-dissipative.

As a direct consequence of nonlocality,
the thermodynamic potentials are
functionals of the state variables
$\udepl$ and $\dam$, here
understood as fields
\cite{Germain:1983}.
In particular, the internal energy of the rod
is the Lagrangian:
\begin{equation}
\Epot(\eps,\dam,\gamma_i)=
\intLL \elib(\eps,\dam) \dd x+
\intLL [\gamma_1\,g_1(\dam)+\gamma_2\,g_2(\dam)]\dd x
\label{eq-Epotential}
\end{equation}
where $\elib$ is the local stored energy function defined by
equation (\ref{eq-freeenergy});
likewise, a global pseudo-potential
of dissipation is obtained by integrating
the dissipation function (\ref{eq-localdiss})
over the physical domain $\Omega$:
\begin{equation}
\Diss(\ddom)=
\intLL\varphi(\ddam) \dd x
\label{eq-Dpotential}
\end{equation}

The forces work-conjugate to the axial strain
and damage are the Cauchy stress
and the energy release rate.
In particular, the former is obtained
along with the equilibrium equation
by zeroing the first variation of the potential energy (\ref{eq-Epotential})
with respect to the displacement $\udepl$ as
 \begin{equation}
\sig=\omegad(\dam)\,E\,\udeplprime
\label{eq-stress}
\end{equation}

For the ensuing developments
the gradient constraint is taken as:
 \be{
 \gdue=\left\vert\damprime\right\vert-\inv{\lc} \leq0
 \label{oned-constraint-lin}}
whence results a piece-wise
linear distribution of damage along the rod.
The energy release rate is a variational derivative
and includes a nonlocal term originating from (\ref{oned-constraint-lin})
plus two boundary conditions \cite{Miehe:2010a}.
In particular, the first variation of the functional
(\ref{eq-Epotential}) with respect to damage
followed by integration by parts yields:
 \begin{equation}
 \frac{\partial\Epot}{\partial\dam}\deldom^*
 =\left[-\intLL G\,\dd x+
 \leftj\gamma_2\,\sign\left(\damprime\right)\rightj_S+ \left[\gamma_2\,\sign\left(\damprime\right)\right]_{\partial\Omega}
 \right]\,\deldom^*
 \label{eq-damagefirstvar-uno}
  \end{equation}
where the domain term $G$ reads:
 \begin{equation}
G=-\frac{\partial\elib}{\partial\dam}-\gamma_1\,\frac{\dd g_1}{\dd\dam}+\frac{\dd }{\dd x}
\bigg(\gamma_2\,\sign\left(\damprime\right)\bigg)
 \label{eq-G-err}
  \end{equation}
$S$ being the set of possible discontinuity points
for the damage gradient.
For non-dissipative internal discontinuities
the (internal) jump relationships give:
 \begin{equation}
\gamma_2^+=\gamma_2^-=0
 \label{eq-gammadue-int-bc}
  \end{equation}
whereas the (external) natural boundary
conditions read:
 \begin{equation}
\gamma_2(x)=0;\quad x\in\partial\Omega
 \label{eq-gammadue-ext-bc}
 \end{equation}
which allow for non-zero damage derivatives
on the outer boundary.

Damage evolution is governed by
the normality rule:
\begin{equation}
G-\Ycd \leq 0, \qquad \ddom\geq 0, \qquad (G-\Ycd)\, \ddom =0
\label{eq-Normality-nonlocal}
\end{equation}
that follows from the Biot-like
subdifferential inclusion:
 \begin{equation}
-\deri{\Epot}{\dom}\in\partial\Diss(\ddom)
\label{eq-biotlike-nonlocal}
\end{equation}


During initial loading
the displacement
$\ustar$ increases
up to the elastic limit:
\be{\uel=\sqrt{\frac{2\,\Yczero}{-\omega^\prime(0)\,E}}\,L
\label{oned-elastlim}}
and the unique response is the
homogeneous elastic one:
\be{
\ustar = \frac{\sigma\,L}{E}
\label{oned-homog}}

Once damage has started to grow and its spatial distribution $\dom(x)$
is known, the relationship between the constant
stress and the prescribed displacement
can be made explicit as:
\be{
\ustar = \frac{\sigma}{2\,E}
\intLL\omegad^{-1}(\dom(x))\dx
\label{oned-Global}}

Solutions beyond the elastic limit
are associated with the initiation
and growth of defects and
can be either homogeneous or localized;
in both cases one can use a parametrization
in terms of the maximum damage level $\dm\leq1$.

For the homogeneous inelastic case, damage evolution
requires the local strain energy
to increase everywhere in the bar;
this can be expressed as
\be{
\left(\frac{\Ycd}{-\omega^\prime(\dom)}\right)^\prime>0
\label{oned-strainenergy}}
where use is made of the local limit condition
emanating from (\ref{eq-Normality-nonlocal}) and
the prime denote differentiation with
respect to the driving variable $\dom$.
The above inequality is equivalent to:
\be{
\Ycd\,\omega^{\prime\prime}(\dom)-\Ycpd\omega^\prime(\dom)>0
\label{oned-stability}}
that is a necessary requirement for local stability.
Condition (\ref{oned-stability}) is suggested
in \cite{LorentzGodard:2011} along with
an additional strain softening condition, whereby
the complementary elastic energy
should decrease with damage, that is:
\be{\left(
\frac{\omegad^2(\dom)\,\Ycd}{-\omega^\prime(\dom)}
\right)^\prime<0
\label{oned-compl-energy}}
whereby one has
\be{
\left[\Ycpd\omegad^2(\dom)+
\Ycd\,2\,\omegad(\dom)\omega^\prime(\dom)\right]\omega^\prime(\dom)-
\Ycd\omegad^2(\dom)\,\omega^{\prime\prime}(\dom)>0
\label{oned-softening-cond}}

%
Without loss of generality, for non-homogeneous
damage we assume that strain localization
associated with one single defect initiates at point
$x=0$ immediately after the
initial elastic limit (\ref{oned-elastlim})
has been attained;
the study can therefore be limited to half of the bar
on account of the symmetry
of (\ref{oned-constraint-lin}).

\begin{figure}[hbt!]
\begin{center}
\includegraphics[width=0.5\textwidth]{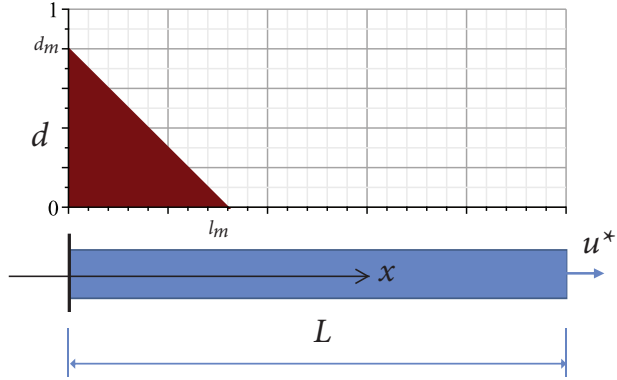}
\end{center}
\caption{One-dimensional rod. Damage distribution for
localized solution.}
\label{fig-oned-2}
\end{figure}

The complementarity conditions (\ref{oned-KTconditions})
imply that the multiplier
$\gamma_2$ can be non-zero
only where the nonlocal constraint
(\ref{oned-constraint-lin})
is met with the equality.
In this case the damage field reads:
\be{ \dom(x) =\dom(0)-\dfrac{x}{\lc}\label{oned-Damage}}
and the constraint set coincides with the interval
$[0,\lm]$, being
\be{\lm =\lc\,\dom(0)=\lc \,\dom_m\leq\lc
\label{oned-Gamma}}
the half-width of the localization band,
see also Figure \ref{fig-oned-2}.

For damage evolution $(\dot\dom>0)$
one has from (\ref{eq-Normality-nonlocal}) the differential problem:
\be{
\Y(\dam(x))
-\frac{\dd\gamma_2}{\dd x}
=\Yc(\dam(x))
\label{oned-eqdiff}}
where the local damage-driving force reads:
\be{
\Y
=-\frac{\partial\elib}{\partial\dam}
=\frac{-\omega^\prime(d)}{\omega^2(d)}
\frac{\sigma^2}{2\,E}
\label{oned-localY}}

Relationship (\ref{oned-eqdiff})
is a first-order differential equation
subject to two boundary conditions; the first one
allows to compute the (uniform) stress $\sigma$
as a function of the driving variable $\dm$
and the second one
is needed to set the integration constant for the
Lagrange multiplier field $\gamma_2$.
The latter is certainly nihil either on
the boundary of the active constraint set
defined by $\gdue=0$,
either where the
gradient of damage is discontinuous, or on the outer
boundary of the domain, where
condition (\ref{eq-gammadue-ext-bc}) holds.
Therefore, integration of (\ref{oned-eqdiff})
between 0 and $\lm$, which correspond to
two discontinuity points for the damage gradient,
provides the \emph{averaged limit condition} \cite{Valoroso-Stolz:2022}:
\be{
\int_{0}^{\lm}\Y(\dam(x))\dd{x}=
\int_{0}^{\lm}\Yc(\dam(x))\dd{x}
}
\label{oned-averaged-limcond}

The integrals are computed via \emph{u}-substitution in the form
\be{
\int_{0}^{\lm}y(\dom(x))\dx=
-\lc\,\int_{\dom(0)}^{\dom(\lm)}y(\dom)\dd\dom
\label{oned-usubstitution}}
and one obtains the stress as a function
of the maximum damage level $\dm$ as:
\be{
\sigma(\dm)=\left[\frac{2\,E}{\omegad^{-1}(\dm)-1}\,\Hm\right]^{\frac{1}{2}}
\label{oned-Tredici}}
where $\Hm$ is the definite integral:
\be{\Hm=
\int_{0}^{\dm}\Ycd\dd\dom
\label{oned-Quattordici}}
and the integrand $\Ycd$ is the
constitutive function, which can
in turn be determined in a way
consistent with a cohesive model.
To this end re-write equation (\ref{oned-Global})
for half of the bar and split the integral
into two parts, respectively accounting for damage behaviour and
a purely elastic response:
\be{
\ustar = \frac{\sigma}{E}
\left[
\int_{0}^{\lm}\left[\omegad^{-1}(\dom(x))-1\right]\dx
+L\right] = \demi\,\wcoh + \frac{\sigma\,L}{E}
\label{oned-Global-cohesive}}

In the above equation $\wcoh$ is the apparent opening
displacement across the localization band;
it can be expressed in terms of the
chosen parametrization as:
\be{
\wcoh(\dm)=\frac{2\,\sigma(\dm)}{E}\,\lc\,\Fm
\label{oned-wcoh}}
where the non-dimensional term $\Fm$ reads:
\be{\Fm=
\int_{0}^{\dm}
\left(\omegad^{-1}(\dom)-1\right)
\dd\dom
\label{oned-Fm}}
and depends only
upon the assumed form
of the degradation function $\omegad(\dom)$.

\begin{figure}[hbt!]
\begin{center}
\includegraphics[width=0.4\textwidth]{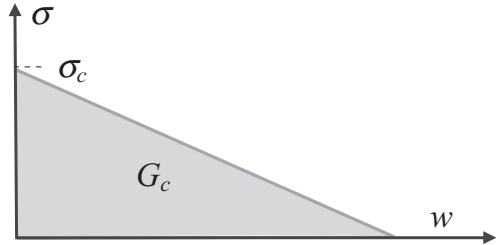}
\end{center}
\caption{Linear softening function.}
\label{fig-oned-3}
\end{figure}

Relationships (\ref{oned-Tredici}) and (\ref{oned-wcoh})
are used to determine the constitutive function $\Ycd$
by requiring that the macroscopic response of the
damageable rod be equivalent to that of an elastic
bar in which the localization band is replaced
by a cohesive interface of given properties.
In particular, we consider the linear softening
law depicted in Figure \ref{fig-oned-3},
whose analytical expression reads:
\be{\sigma
=\sigmaf\left(
1-\frac{\sigmaf}{2\,\Gf}\,\wcoh
\right)
\label{oned-linsoft}}
where $\sigmaf$ and $\Gf$
respectively denote the peak stress
and the fracture energy.

%
Substitution of (\ref{oned-wcoh}) into (\ref{oned-linsoft})
and solution for $\sigma(\dm)$
provides the stress as
a function of the maximum damage $\dm$ as:
\be{\sigma(\dm)=
\frac{\sigmaf}{1+\lambda\,\Fm}
\label{oned-sigma-coh}}
where $\lambda$ is the non-dimensional parameter:
\be{
\lambda = \frac{\lc}{\lcoh}
\label{oned-lambda}}
expressing the ratio between the characteristic length $\lc$
of the graded damage model and
the length scale $\lcoh$ of the cohesive zone \cite{Hillerborg76}:
\be{
\lcoh=\frac{E\,\Gf}{\sigmafq}
\label{oned-lambda-lcoh}}

%
For a given $\dm$, the
value  $\Hm$ of the integral (\ref{oned-Quattordici})
follows from substitution
of (\ref{oned-sigma-coh})
into (\ref{oned-Tredici}) as:
\be{\Hm=
\frac{\omegad^{-1}(\dm)-1}{2\,E}
\left(\frac{\sigmaf}{1+\lambda\,\Fm}\right)^2
\label{oned-Hm-sol}}

Motivated by stability arguments
that are being illustrated later on,
we choose for the degradation function
the quadratic expression:
\be{
\omegad=(1-\dom)^2
\label{oned-omegad}}
whereby one has from (\ref{oned-Fm}):
\be{
\Fm=\frac{\dm^2}{1-\dm}
\label{oned-Fm-sol}}
while (\ref{oned-Tredici}) provides
the following expression for the (uniform) stress:
\be{\sigma(\dm)=
\sigmaf\,\frac{1-\dm}{\lambda\,\dm^2+1-\dm}
\label{oned-sigmadm-sol}}
with the limits
\be{
\lim_{\dm\rightarrow0}\sigma(\dm)=\sigmaf;
\qquad
\lim_{\dm\rightarrow1}\sigma(\dm)=0
\label{oned-sigmadm-limit}}

The constitutive function $\Ycd$ is computed
by differentiation of (\ref{oned-Hm-sol}) as:
\be{
\Ycd = \frac{\sigmafq}{E}\;\frac{1+\lambda\,\dom^2(\dom-3)}
{\left( \lambda\,\dom^2+1-\dom\right)^3}
\label{oned-Ycd-sol}}
and the relevant limits read:
\be{
\Yczero=\lim_{\dom\rightarrow0}\Ycd=\frac{\sigmafq}{E};
\qquad
\lim_{\dom\rightarrow1}\Ycd=\frac{\sigmafq}{E}\;\frac{1-2\,\lambda}{\lambda^3}
\label{oned-Ycd-limit}}

\begin{figure}[hbt!]
\begin{center}
\includegraphics[width=0.4\textwidth]{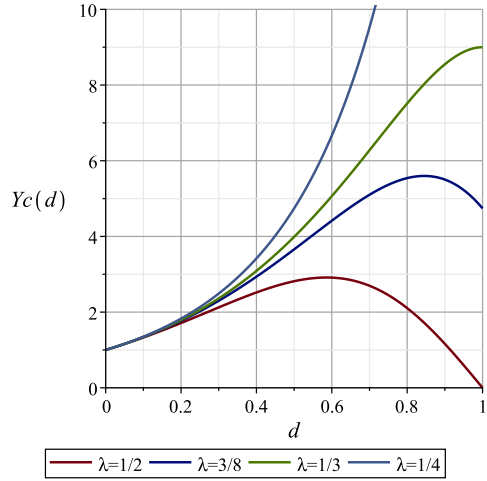}
\end{center}
\caption{The constitutive function $\Yc$
consistent with linear softening.}
\label{fig-oned-4}
\end{figure}

Figure \ref{fig-oned-4}
depicts the function $\Ycd$
normalized to $\Yczero$
for different values of the parameter
$\lambda$, which determines the
properties of the function $\Ycd$ itself.
In practice, a \emph{safe value} of $\lambda$ to be used
in numerical computations \cite{Valoroso-Stolz:2022}
can be taken in a way to comply
with conditions (\ref{oned-stability})
and (\ref{oned-softening-cond}).

For the case at hand the local stability
requirement (\ref{oned-stability}) provides:
\be{
\lambda<
{\frac { \left( \dom-2 \right) \sqrt {{\dom}^{4}-4\,{\dom}
^{3}+40\,{\dom}^{2}-72\,\dom+36}+{\dom}^{3}-4\,{\dom}^
{2}-10\,\dom+12}{8\,{\dom}^{4}-36\,{\dom}^{3}+24\,
{\dom}^{2}}}
\label{oned-stab-lambda}}
with limit
\be{
\lim_{\dom\rightarrow 1}\lambda=\demi
\label{oned-lambda-limit}}

On the other hand, the strain softening
condition (\ref{oned-softening-cond}) implies:
\be{
0\,<\,\lambda\,<\,\frac{1+(1-\dom)^2}{2\,\dom}
\label{oned-lambda-soft}}

\begin{figure}[hbt!]
\begin{center}
\includegraphics[width=0.4\textwidth]{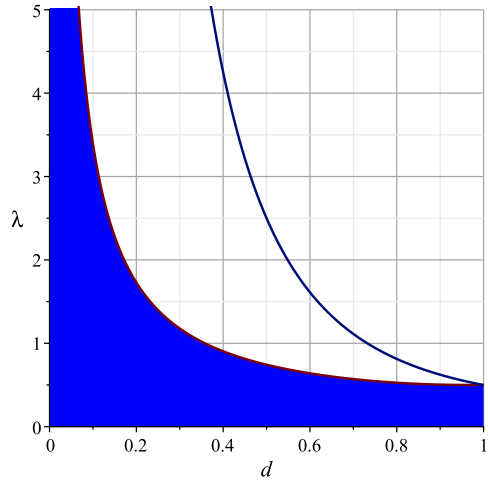}
\end{center}
\caption{Admissible region for parameter $\lambda$.}
\label{fig-oned-5}
\end{figure}

The above relationships
define the admissible region for $\lambda$
that is shaded in Figure \ref{fig-oned-5}.
Clearly, any positive value of $\lambda$ lower than $0.5$
allows to fulfill both conditions for
each $\dom\,\in\,[0,\,1]$.

For a given damage
distribution the relationship
between the stress and the
displacement $\ustar$ is obtained
from equation (\ref{oned-Global}) as:
\be{
\sigma(\dm)=\frac{E\,\ustar}{L}\;
\frac{1-\dm}{\bigl( \beta\,\dm^2+1-\dm \bigr)}
\label{oned-Tredicib}}
with
\be{
\beta=\frac{\lc}{L}
\label{oned-beta}}

Evidently, in the present context
knowledge of the damage distribution
is equivalent to knowledge of
the constraint set $[0,\lm]$, where
the integral in equation (\ref{oned-Global}) is non-trivial
and the Lagrange multiplier $\gamma_2$ is non-zero.

Evaluation of the multiplier $\gamma_2$
amounts to compute the integral
of the differential equation (\ref{oned-eqdiff}).
To this end use is made of the chain rule as:
\be{
\dfrac{\dd\gamma_2}{\dx}=
\gamma_2^\prime(\dom)\dfrac{\dd\dom}{\dx}
=-\inv{\lc}\gamma_2^\prime(\dom)
\label{oned-Dodici}}
to get the integral as:
\be{
\frac{1}{\lc}\,\gamma_2(d)=\frac{-\sigma^2(\dm)}{2\,E\,\omegad(\dom)}
+
\frac{\sigmafq}{E}\;\frac{\dom\,(2-\dom)}
{2\,\left( \lambda\,\dom^2+1-\dom\right)^2}+C
\label{oned-Quattordicib}}

The integration constant $C$ is obtained
using one of the two boundary conditions on $\gamma_2$,
i.e.~$\gamma_2(0)=\gamma_2(\dom_m)=0$, that is:
\be{
C=\frac{\sigma^2(\dm)}{2\,E}
\label{oned-Quindicib}}
whereby one has:
\be{
\begin{array}{ll}
  \gamma_2(d)&=
  \displaystyle{\frac{-\sigma^2(\dm)\,\lc}{2\,E}\left( \omegad^{-1}(\dom)-1\right)+
\frac{\sigmafq\,\lc}{E}\;\frac{\dom\,(2-\dom)}{2\,\left( \lambda\,\dom^2+1-\dom\right)^2}}
 \\
 \noalign{\bigskip}
  &=
  \displaystyle{\frac{\lc}{2\,E}\,\frac{\dom\,(2-\dom)}{(1-\dom)^2}
  \bigl[\sigma^2(\dom)-\sigma^2(\dm)\bigr]}
\end{array}
\label{oned-Sedicib}}

Equation (\ref{oned-Sedicib})
describes the variation of the multiplier $\gamma_2$
within the active constraint interval and
depends upon $\dm$, which is a fixed value,
and $\dom=\dom(x)$, which is a function
of the abscissa $x$ along the bar.
It is immediately recognized that
the Lagrange multiplier $\gamma_2$ is zero either
at $x=0$, where $\dom=\dm$,
and at $x=\lm$, where $\dom=0$.
Moreover, its
maximum value
within the interval occurs at point $\bar{x}$ where
$\Y(x)=\Yc(\dam(x))$
owing to the differential relationship (\ref{oned-eqdiff}).

\begin{figure}[hbt!]
\begin{center}
\includegraphics[width=0.4\textwidth]{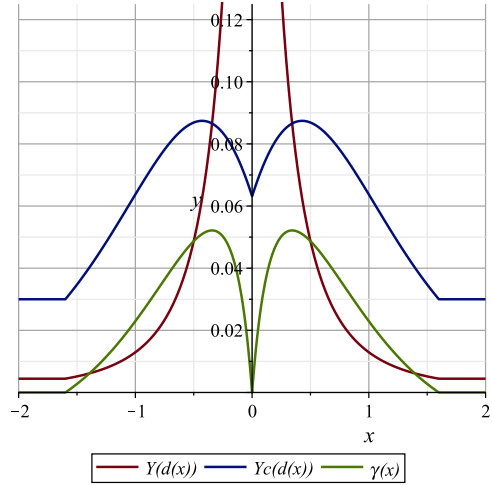}
\end{center}
\caption{One-dimensional rod. A typical spatial distribution of $\Y$, $\Yc$ and
$\gamma_2$ corresponding to the symmetric localized solution with a single defect.}
\label{fig-oned-6}
\end{figure}

A typical spatial distribution of the Lagrange multiplier $\gamma_2(x)$
within the symmetric localization band is depicted in Figure
\ref{fig-oned-6} along with the local damage-driving force
$\Y$ and the constitutive
function $\Yc$.

The response of the elasto-damaging bar
is clearly dependent upon the length scales $\lc$
and $\lcoh$ via the non-dimensional
parameter $\lambda$ defined by (\ref{oned-lambda})
and from the geometric factor $\beta$
given by (\ref{oned-beta}).

Actually, by comparison of equation (\ref{oned-sigmadm-sol})
with (\ref{oned-Tredicib}) one has:
\be{
\ustar(\dm)=\frac{\sigmaf\,L}{E}\;
\frac{\beta\,\dm^2+1-\dm}
{\lambda\,\dm^2+1-\dm}
\label{oned-uubeta}}

This relationship allows to
obtain the condition
under which the response of the rod
is stable under displacement control,
i.e.~it does not exhibit a snap-back.
This requires the end-displacement $\ustar$
to be an increasing function of the
maximum damage level $\dm$, that is:
\be{
\frac{\partial\ustar}{\partial\dm}=\frac{\sigmaf\,L}{E}\;
\frac{(\beta-\lambda)(2\,\dm-\dm^2)}
{\lambda\,\dm^2+1-\dm}>0
\label{oned-uubeta-derivative}}
whereby one obtains the condition
that governs the stability of the response
for the damaging tensile bar
under displacement control
\be{
\beta>\lambda \;\Leftrightarrow\;
L<\lcoh
\label{oned-uubeta-condition}}

The stability condition strongly depends upon the
expressions of the degradation function (\ref{oned-omegad})
and of the constitutive function
(\ref{oned-Ycd-sol}).
Actually, taking for $\Yc$ the constant
function, i.e.~$\Yczero$ given by (\ref{oned-Ycd-limit}), and
the quadratic degradation function (\ref{oned-omegad})
one obtains:
\be{
\sigma(\dm)=\frac{2\,\sigmaf\,(1-\dm)}{\sqrt{4-2\,\dm}}
\label{oned-sigmadm-sol-const}}
in place of (\ref{oned-sigmadm-sol}) and
\be{
\ustar(\dm)=\frac{\sigmaf\,L}{E}\;
\frac{2(\beta\,\dm^2+1-\dm)}
{\sqrt{4-2\,\dm}}
\label{oned-uubeta-const}}
that replaces (\ref{oned-uubeta}).
The stability condition now reads:
\be{
\beta>\frac{3-\dm}{\dm\,(8-3\,\dm)}
\label{oned-uubeta-const-condition}}
whereby one infers that there is always a snap back
right after the elastic limit
no matter how short is the bar
since the right-hand side of (\ref{oned-uubeta-const-condition})
diverges for $\dm\rightarrow0$.
This can slow down convergence
in the solution of a Finite Element
problem and should be avoided as much as possible.

However, there exist situations that are even more
harmful. In this respect, consider
the case of a linear degradation function
\be{
\overline{\omegad}(\dom) = 1-\dom
\label{oned-omegad-umd}}
and a constant elastic limit
\be{
\overline{\Y}_c = \frac{\sigmaf^2}{2\,E}
\label{oned-Yc-umd}}

In this case one obtains the stress and the end displacement as:
\be{
\sigma(\dm)=\sigmaf\,\sqrt{1-\dm}
\label{oned-sigmadm-unomd}}
\be{
\ustar(\dm)=\frac{\sigmaf\,L}{E}\;
\bigl(
1-\beta\,\dm-\beta\,\ln{(1-\dm)}
\bigr)
\label{oned-uubeta-c-unomd}}
The stability condition now reads:
\be{
\beta>\frac{1}{\ln{(1-\dm)}+3\,\dm}
\label{oned-uubeta-c-unomd-condition}}

Clearly, the right-hand side
of (\ref{oned-uubeta-c-unomd-condition})
diverges for both $\dm\rightarrow0$ and
for $\dm\rightarrow 1-exp(-3\dm)\simeq0.94048$.

\begin{figure}[hbt!]
\begin{center}
\includegraphics[width=0.5\textwidth]{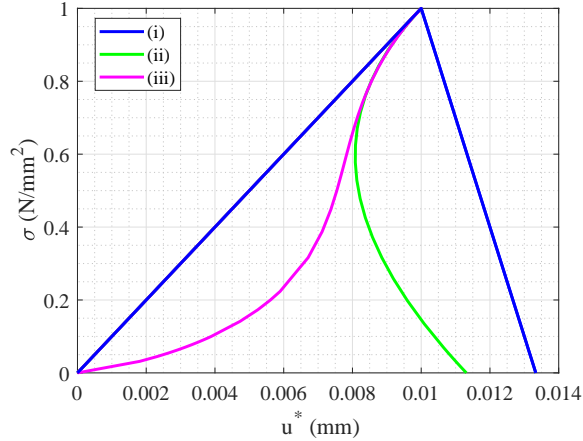}
\end{center}
\caption{One dimensional rod.
Normalized stress-displacement responses obtained
for different choices of the constitutive
functions $\omegad(\dam)$ and $\Ycd$.}
\label{fig-oned-7}
\end{figure}

Figure \ref{fig-oned-7} depicts the
$\sigma-\udepl^\star$ response of the bar
for the different choices of the constitutive functions
$\omegad(\dom)$ and $\Ycd$
considered above, that is:
\begin{enumerate}
  \item[(i)] quadratic degradation and non-constant limit $\Ycd$ that
realizes the equivalence with linear softening;
  \item[(ii)] quadratic degradation and constant limit $\Yc=\sigmaf^2/E$;
  \item[(iii)] linear degradation and constant limit $\Yc=\sigmaf^2/{2E}$.
\end{enumerate}
For all these cases
the non-dimensional parameters
respectively defined by (\ref{oned-lambda}) and (\ref{oned-beta})
are such that $\lambda\leq0.5$ and $\beta>\lambda$,
which correspond to a rod that can be considered a short one.


\section{The block with cohesive interface}
\label{sec3}

Generally speaking, a cohesive law
is a relationship between a displacement discontinuity vector,
which is understood as the interface strain,
and a surface traction vector playing the role of the stress.
For the developments that follow
attention will be restricted to mode-I opening;
tractions and displacement jumps will then be
normal to the interface
while negative relative displacements
will be left out for notational simplicity.

\begin{figure}[hbt!]
\begin{center}
\includegraphics[width=0.4\textwidth]{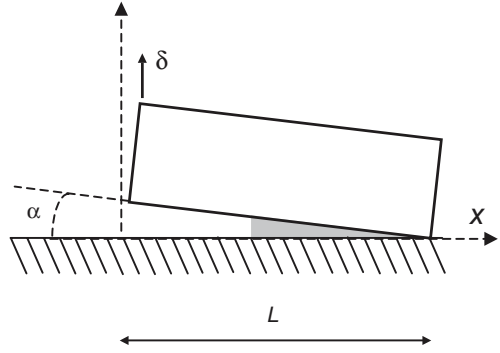}
\end{center}
\caption{The rigid-block problem.}
\label{fig-8-block}
\end{figure}

As a model problem consider the
structure in Figure \ref{fig-8-block},
consisting of a rigid block
connected to a fixed support via
a damageable adhesive layer of negligible thickness \cite{Volokh:2004}.
A monotonic increase
of the end-displacement $\delta$ produces
a uniform rotation of the block but
a non-uniform distribution of damage,
which starts nucleating from the left edge
with non-zero gradient.

Denoting by $\epsint$
the opening displacement across the interface,
a stored energy function
from which one can obtain a (local) cohesive law
using a damage-based formulation reads \cite{Valoroso-06}:
\begin{equation}
\elibint(\epsint,\dam)=\unmezzo\,\omegad(\dam)\,\kk\epsintq
\label{eq-freeenergy-interf}
\end{equation}
where $\kk$ is the (undamaged) interface stiffness
in tension.

For the problem at hand
the kinematics of deformation is completely
described by a single parameter, i.e.~the
rotation $\alpha$,
here assumed to be small in the usual sense.
Therefore, one has the opening displacement:
\begin{equation}
\epsint(x)=\alpha\,(L-x)
\label{eq-disp-field}
\end{equation}
while the stress-like variables read:
\begin{equation}
\begin{array}{c}
 \sigint(x,\dam) =\displaystyle{\frac{\partial\elibint(\epsint,\dam)}{\partial\epsint}
 =\omegad(\dam)\,\kk\alpha\,(L-x)}\\
\noalign{\bigskip}
\displaystyle{
  \Yint(x,\dam)=\displaystyle{-\frac{\partial\elibint(\epsint,\dam)}{\partial\dam}
  =-\omegad^\prime(\dam)\,\unmezzo\,\kk\alpha^2(L-x)^2}
  }
\end{array}
\label{eq-block-forces}
\end{equation}

The governing equations of a nonlocal interface model
based on the graded damage concept
are formally identical to those
developed in Section \ref{sec2} for the tensile rod
with two minor modifications.
In particular, the stored energy function
(\ref{eq-freeenergy}) has to be replaced by
(\ref{eq-freeenergy-interf}), while
the function $\Ycdint$
is now directly prescribed,
e.g.~based on the shape of a chosen
traction-separation relationship.

\begin{figure}[hbt!]
\begin{center}
\includegraphics[width=0.4\textwidth]{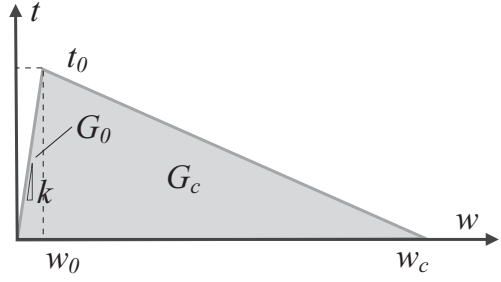}
\end{center}
\caption{Bilinear cohesive law}
\label{fig-9-bilinear}
\end{figure}

For instance,
a constitutive function $\Ycdint$
that yields the (local) bilinear cohesive law
of Figure \ref{fig-9-bilinear} reads:
\begin{equation}
\Ycdint=\frac{-\omegad^\prime(\dam)\,\Gzero\,\Gc^2}
{\big[ \Gzero+\left(\Gc-\Gzero\right)\,\omegad(\dam)\big]^2}
\label{eq-Yc}
\end{equation}
to which corresponds the work of separation:
\begin{equation}
\int_{0}^{+\infty}\Ycdint\,\ddam\,\dd t=
-\int_{0}^{1}\Ycint(\omegad)\,\dd \omegad=
\Gc
\label{eq-Yc-integral}
\end{equation}
where $\Gzero$ and $\Gc$
respectively denote the initial energy threshold
and the interface fracture toughness:
\begin{equation}
\Gzero=\unmezzo\kk\epsintzero^2; \qquad
\Gc=\unmezzo\kk\epsintzero\,\epsintc
\label{eq-Gzero-Gc}
\end{equation}

Without loss of generality,
in the remainder we shall
assume equations (\ref{oned-constraint-lin})
and (\ref{oned-omegad}) to hold;
moreover, the length scale $\lc$
of the nonlocal interface model is supposed
to be greater than the width $L$ of the block.

The equilibrium path of the structure
can be traced using the balance of moments
about the center of rotation:
\begin{equation}
\Preac\,L = \intL\sigint(x,\dam)\,(L-x)\dd x
\label{eq-balance-momts}
\end{equation}
$\Omega$ being the physical domain
$[0,L]$ and
$\Preac$ the reaction force
corresponding to the prescribed displacement $\delta$.
During initial loading the latter increases
up to the elastic limit $\delta_0$ given by:
\begin{equation}
\delta_0
=\sqrt{\frac{2\,\Gzero}{\kk}}
=\alpha_0\,L
\label{eq-deltazero}
\end{equation}
that is first attained
when the local limit condition
$\Yint=\Ycint(0)$ is met.
This state corresponds to
damage nucleation at $x=0$ and
from this point onwards loading can be
effectively parametrized in terms of the
size $\lm>0$ of the damaged portion
of the domain of interest.

For $\lm\leq\lc$ the
length $\lm$ does also coincide with the size of the
active constraint set (i.e.~where $g_2(\dam)=0$)
and the adopted parametrization
is fully equivalent to the one given
in terms of the maximum damage $\dm$
defined as:
\begin{equation}
\dm=\min\left\{
1, \frac{\lm}{\lc}
\right\}
\label{eq-dm-definition}
\end{equation}

It is worth emphasizing
that, unlike the case of the
tensile rod with a single evolving defect,
for the problem at hand the constraint set
translates along the interface once
the damage process zone has fully developed.
In particular, this occurs when the size of the
damaged region $\lm$ equals the length scale $\lc$;
to account
for this case, the (piece-wise) linear damage function
that is prescribed via the gradient constraint (\ref{oned-constraint-lin})
is conveniently defined as:
\begin{equation}
\dom(x) =\max\left\{0, \dm-\frac{x-c}{\lc}\right\}
\label{eq-damage-distrib-interface}
\end{equation}
where the (finite)
size of the fully damaged subdomain reads:
\begin{equation}
c=\max\bigg\{0, \lm-\lc \bigg\}
\label{eq-damage-front-c}
\end{equation}

As discussed in Section \ref{sec2},
owing to gradient-dependence
the normality rule yields the
differential equation (\ref{oned-eqdiff});
the latter now admits two sets of boundary conditions
for the opening angle $\alpha$
and the Lagrange multiplier $\gamma_2$.

For the rigid block problem
the averaged limit condition reads:
\begin{equation}
\int_{0}^{H}\Yint(x,\dam(x))\dd x
=\int_{0}^{H}\Ycint(\dam(x)) \dd x
\label{eq-averaged-overH}
\end{equation}
where $H$ denotes the size of the
active process zone portion that is
contained within the physical domain $[0,L]$:
\begin{equation}
H=\min\bigg\{\lm, L \bigg\}
\label{eq-activeproczone}
\end{equation}

With this notation in hand,
balance of moments about the center of rotation can be expressed as:
\begin{equation}
\Preac\,L=
\int_{c}^{H}\omegad(\dam)\kk\hat{\alpha}\,(L-x)^2\dd x
+\int_{H}^{L}\kk\hat{\alpha}\,(L-x)^2\dd x
\label{eq-Preac-nonlocal}
\end{equation}
whereby one obtains the different branches of the
equilibrium path by
distinguishing
the different possible cases
for the integration limits
of equation (\ref{eq-averaged-overH}),
which in turn take into account
the boundary conditions (\ref{eq-gammadue-ext-bc})
for the Lagrange multiplier field
and the internal jump conditions
(\ref{eq-gammadue-int-bc}), if any.

\begin{figure}[hbt!]
\begin{center}
\includegraphics[width=0.5\textwidth]{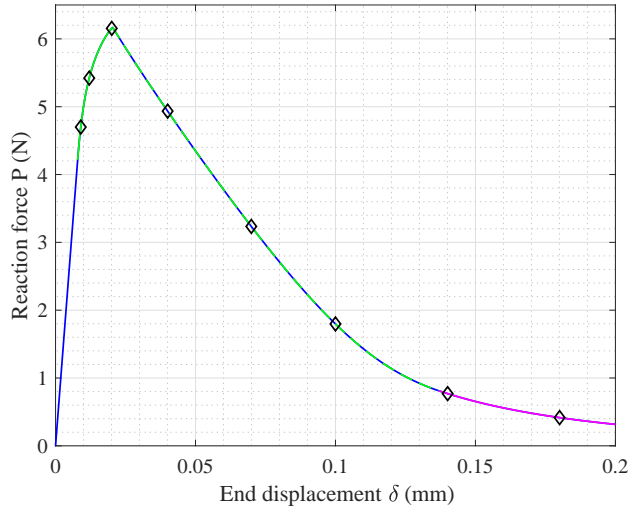}
\end{center}
\caption{Equilibrium curve for the rigid-block problem.}
\label{fig-10-nonlocal-curve}
\end{figure}

\begin{table}
  \caption{Data set for the rigid block problem}
  \label{tab-uno}
  \centering
\begin{tabular}{|c|c|}
\hline
  $L=2\,mm$; & $\kk=800\,N/mm^3$   \\
\hline
  $\Gc=0.25\,N/mm$; & $\Gzero=0.025\,N/mm$  \\
\hline
\end{tabular}
\end{table}


\paragraph{Phase 1. Linear elastic}

The initial linear elastic
phase is purely local;
the limit value for the
displacement is given by
(\ref{eq-deltazero}),
to which corresponds the reaction force:
\begin{equation}
\Preac_0
=\sqrt{2\,\Gzero\,\kk}\,\frac{L}{3}
\label{eq-Pzero}
\end{equation}
%

\paragraph{Phase 2. Damage nucleation and growth for $0<\lm\leq{L}$}

In this case use of the integral limit condition (\ref{eq-averaged-overH})
allows to compute the opening
angle $\alpha$ as a function of the length $\lm$
(driving variable) as:
\begin{equation}
\hat{\alpha}_1=\alpha_0\,\sqrt{\frac{\Gc\,\lc^2\,L^2\,(2\,\lc-\lm)}{\Auno\,\Adue}}
\label{eq-alpha1-nonlocal}
\end{equation}
with
\begin{equation}
\Auno = \left[\frac{\lm^3}{6}-\frac{2}{3}(\lc+L)\,\lm^2
+(L+2\,\lc)\,L\,\lm - 2\,L^2\,\lc \right]
\label{eq-alpha1-nonlocal-Auno}
\end{equation}
\begin{equation}
\Adue = \left[ (\Gzero-\Gc)(\lm^2-2\,\lc\,\lm)-\Gc\,\lc^2 \right]
\label{eq-alpha1-nonlocal-Adue}
\end{equation}
and limits
\begin{equation}
\hat{\alpha}_{1}^0= \lim_{\lm\rightarrow 0}\hat{\alpha}_1=\alpha_0
\label{eq-dieci-vol-lim0}
\end{equation}
\begin{equation}
\hat{\alpha}_{1}^L=\lim_{\lm\rightarrow L}\hat{\alpha}_1=
\alpha_0\;
\sqrt{\frac{6\,\Gc\,\lc^2\,(2\,\lc-L)}
{(4\,\lc-3\,L)\, [(\lc-L)^2\,\Gc-(L-2\,\lc)\,\Gzero\,L]}}
\label{eq-dieci-vol-lim1}
\end{equation}

Obviously, the limit (\ref{eq-dieci-vol-lim0}) coincides with the
opening angle $\alpha_0$ defined by (\ref{eq-deltazero})
whereas the upper limit (\ref{eq-dieci-vol-lim1})
marks the end of the domain of validity of
relationship
(\ref{eq-alpha1-nonlocal}),
to which corresponds by equilibrium
the end reaction force:
\begin{equation}
\hat{\Preac}_1=
\left[
L^3\,\lc^2+(\lm-3\,\lc)\,\lm^2\,L^2-\frac{(\lm-4\,\lc)\,\lm^3\,L}{2}+
\frac{(\lm-5\,\lc)\,\lm^4}{10}
\right] \frac{\kk\hat{\alpha}_1}{3\,\lc^2\,L}
\label{eq-P-phase2-nonlocal}
\end{equation}


\begin{figure}[hbt!]
\begin{center}
\includegraphics[width=0.5\textwidth]{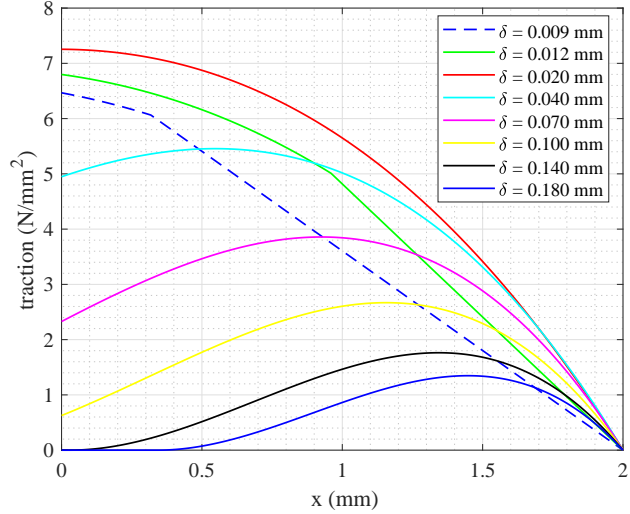}
\end{center}
\caption{Tractions distributions along the interface
at varying end-displacement $\delta$.}
\label{fig-11-nonlocal-curve}
\end{figure}


\paragraph{Phase 3. Damage growth for $L\leq\lm\leq\lc$}

When the driving variable $\lm$
grows beyond the length $L$, the
zero-damage boundary would be located outside the
physical domain of the interface.
In this case the boundary conditions for the Lagrange multiplier
and the limit condition (\ref{eq-averaged-overH})
yield an averaged equality
over the entire domain $[0,L]$,
whereby one obtains another
nonlinear branch of the equilibrium path
defined by the following:
%
\begin{equation}
\hat{\alpha}_2 = \alpha_0\,
\sqrt{\frac{6\,\Gc^2\,\lc^4\,(L+2\,\lc-2\,\lm)}{B_1\,B_2\,B_3}}
\label{eq-alpha2-nonlocal}
\end{equation}
with
\begin{equation}
B_1=L+4\,\lc-4\,\lm
\label{eq-alpha2-nonlocal-B1}
\end{equation}
\begin{equation}
B_2=(\lc-\lm)^2\,\Gc-\lm\,(\lm-2\,\lc)\,\Gzero
\label{eq-alpha2-nonlocal-B2}
\end{equation}
\begin{equation}
B_3=(L-\lm+\lc)^2\,\Gc-
(L-\lm)(L-\lm+2\,\lc)\Gzero
\label{eq-alpha2-nonlocal-B3}
\end{equation}

The corresponding value of the reaction force reads:
\begin{equation}
\hat{\Preac}_2=
\frac{\kk\,L^2}{30\,\lc^2}
\left[
10\,\lc^2+(5\,L-20\,\lm)\,\lc+L^2-5\,L\,\lm+10\,\lm^2
\right]\,\hat{\alpha}_2
\label{eq-P-phase3-nonlocal}
\end{equation}

\begin{figure}[hbt!]
\begin{center}
\includegraphics[width=0.5\textwidth]{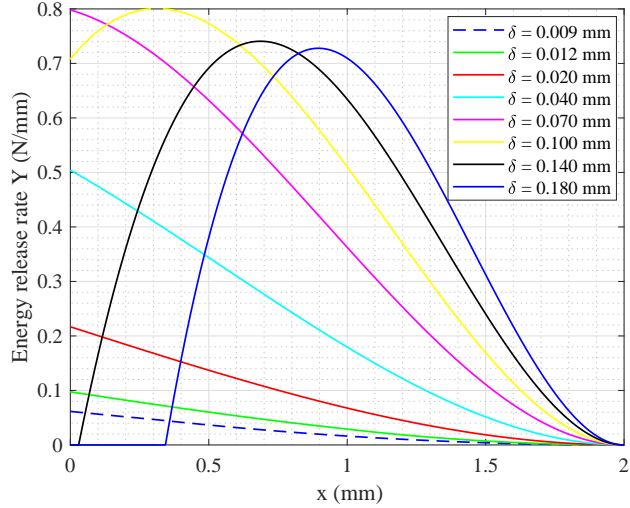}
\end{center}
\caption{Damage-conjugate force $\Yint$ along the interface
at varying end-displacement $\delta$.}
\label{fig-12-nonlocal-curve}
\end{figure}


\paragraph{Phase 4. Crack propagation}

For $\lm>lc$ there exists a fully
damaged region of finite size
$c$ defined by (\ref{eq-damage-front-c});
the latter can be taken as the
driving variable for computing the last part of the
equilibrium curve because in this case the
non-trivial limit condition
reduces to an averaged equality
over the interval $[c,L]$ on account of the
jump relationships (\ref{eq-gammadue-int-bc}).

The opening angle is now computed as:
\begin{equation}
\hat{\alpha}_3=\frac{\alpha_0}{(L-c)^2}
\,\sqrt{
\frac{6\,\Gc^2\,\lc^2\,L^2}
{\Gzero\,\left[\Gc\,(L-c)^2-\Gzero(-L+c-\lc)(-L+c+\lc)\right]}}
\label{eq-alpha3-nonlocal}
\end{equation}
whereas the equilibrium equation
yields:
\begin{equation}
\hat{\Preac}_3=
\frac{\kk\,(L-c)^5}{30\,\lc^2\,L}\,\hat{\alpha}_3
\label{eq-Preac-phase4-nonlocal}
\end{equation}

As expected, the reaction force (\ref{eq-Preac-phase4-nonlocal})
converges to zero when the portion of the active
damage process zone lying within
the physical domain
progressively
shrinks and collapses to a point.

\begin{figure}[hbt!]
\begin{center}
\includegraphics[width=0.5\textwidth]{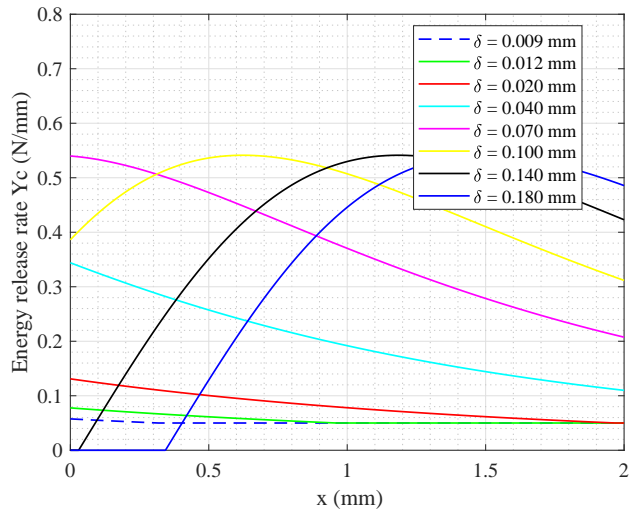}
\end{center}
\caption{Threshold function $\Ycint$ along the interface
at varying end-displacement $\delta$.}
\label{fig-13-nonlocal-curve}
\end{figure}

The complete equilibrium path
for the block delamination problem
is depicted in Figure \ref{fig-10-nonlocal-curve}.
The curve corresponds to a length scale $\lc=6\,mm$
and to the data set of Table \ref{tab-uno}.
The different colors on the plot are used to
distinguish the four branches of the theoretical solution
whereas the points highlighted on the load-deflection
curve indicate the stations selected for the plots of
Figure \ref{fig-11-nonlocal-curve}, which shows the
distribution of the surface tractions along
the interface for different damage levels.

Unlike the case of a local model,
where the profile of the surface tractions
would replicate the bilinear
shape of the traction-separation curve
consequent to (\ref{eq-Yc}),
due to gradient-dependence
the tractions distribution
in the present case changes continuously
from a bilinear shape to an exponential-like one
for increasing end-displacement $\delta$.

Moreover, it is noted that the differential character
of the constitutive relationship
allows for point-wise values of the surface tractions
higher than the peak stress $\sqrt{2\,\kk\,\Gzero}$
of the underlying local model.
Likewise, the point-wise
values of the damage-conjugate variable $\Yint$
can exceed those of the limit function $\Ycint$
owing to the averaged character of the
limit condition. This behaviour
is clearly put forward in
Figures \ref{fig-12-nonlocal-curve}
and \ref{fig-13-nonlocal-curve},
that have been plotted using the same
scale to ease the comparison.

\begin{figure}[hbt!]
\begin{center}
\includegraphics[width=0.5\textwidth]{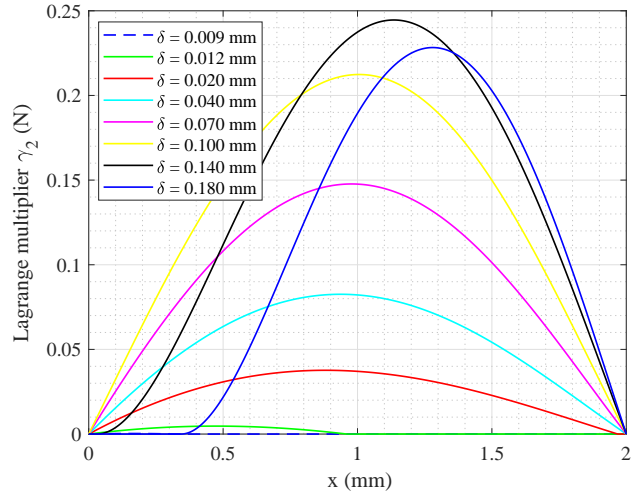}
\end{center}
\caption{Lagrange multiplier $\gamma_2$
along the interface at varying end-displacement $\delta$.}
\label{fig-14-nonlocal-curve}
\end{figure}

It is also worth noting that
functions $\Yint$ and $\Ycint$
do share only one common point-wise
value over the physical domain $[0,L]$.
This corresponds to the extremum
point, which is indeed a maximum
because of the obvious sign constraint,
of the Lagrange multiplier field
$\gamma_2$ that is used to enforce
the nonlocal, gradient constraint,
see e.g.~Figure \ref{fig-14-nonlocal-curve}.

\section{Closure}
\label{sec4}

Based on the graded damage formulation
contributed in \cite{Valoroso-Stolz:2022}
two problems in one dimension have been
discussed and the relevant analytical
solutions computed to be used as a reference
for finite element procedures.
In the tensile rod problem the
uniform stress is obtained from the
averaged limit condition and the
the constitutive function $\Yc$
is determined in a way to produce a global response curve
that is consistent with a cohesive zone model
with linear softening.
On the other hand, in the block delamination
problem the constitutive function $\Yc$
is prescribed based on a local cohesive model
while the integral limit condition
provides the parameter governing the
kinematics of deformation.
In this case one obtains a global equilibrium
curve that depends from the length scale
since the effect of gradient-dependence
is that of relaxing the surface tractions
in that they are no longer
constrained to the shape of the underlying
local softening curve.



\end{document}